\documentclass[aip,prb,twocolumn,showpacs]{revtex4}

\usepackage{graphics,graphicx,amssymb,amsmath,color}
\begin{document}

\title{Paramagnetic spin correlations in colossal magnetoresistive La$_{0.7}$Ca$_{0.3}$MnO$_{3}$}
\author{Joel S. Helton$^{1,\ast}$}
\author{Matthew B. Stone$^{2}$}
\author{Dmitry A. Shulyatev$^{3}$}
\author{Yakov M. Mukovskii$^{3}$}
\author{Jeffrey W. Lynn$^{1,\dag}$}

\affiliation{$^{1}$NIST Center for Neutron Research, National Institute of Standards and Technology, Gaithersburg, Maryland 20899, USA}
\affiliation{$^{2}$Quantum Condensed Matter Division, Oak Ridge National Laboratory, Oak Ridge, Tennessee 37831, USA}
\affiliation{$^{3}$National University of Science and Technology ``MISiS," Moscow 119991, Russia}

\date{\today}
\begin{abstract}
Neutron spectroscopy measurements reveal dynamic spin correlations throughout the Brillouin zone in the colossal magnetoresistive material La$_{0.7}$Ca$_{0.3}$MnO$_{3}$ at 265~K ($\approx$1.03~$T_{C}$).  The long-wavelength behavior is consistent with spin diffusion, yet an additional and unexpected component of the scattering is also observed in low-energy constant-$E$ measurements, which takes the form of ridges of strong quasielastic scattering running along ($H$~0~0) and equivalent directions.  Well-defined $Q$-space correlations are observed in constant-$E$ scans at energies up to at least 28~meV, suggesting robust short-range spin correlations in the paramagnetic phase.

\end{abstract}
\maketitle

Hole-doped perovskite manganites of the form La$_{1-x}$Ca$_{x}$MnO$_{3}$ (LCMO) feature a ferromagnetic metallic ground state for 0.2~$<$~$x$~$<$~0.5, with the highest $T_{C}$ for the combined ferromagnetic and metal-insulator transition at an optimal doping of $x$~$\approx$~3/8.\cite{Wollan,Schiffer,Cheong}  The colossal magnetoresistance (CMR) observed at this transition\cite{Ramirez} cannot be explained solely by Zener double-exchange;\cite{Millis} rather, the physics underlying the CMR effect in LCMO likely arises from strong coupling between the spin, charge, lattice, and orbital degrees of freedom\cite{Roeder,Varma} and nanoscale inhomogeneities between competing phases.\cite{Dagotto,Dagotto2005}  For $x$~=~0.3 LCMO, short-range static and dynamic polaron correlations are observed above $T_{C}$\cite{Lynn2007,Bridges} with an ordering wave vector of (0.25~0.25~0) signifying CE-type\cite{Goodenough} charge- and orbital-ordered regions.  The spin dynamics in the ferromagnetic phase are likewise unconventional.\cite{Lynn2000,Zhang}  The spin wave dispersion softens near the zone boundary, which can be fit to a phenomenological model of first- and fourth-nearest-neighbor ferromagnetic Heisenberg interactions, and displays anomalous spin wave damping.\cite{Ye,Dai}  The spin wave stiffness coefficient renormalizes but does not fully collapse as $T_{C}$ is approached from below;\cite{Lynn1996} this contrasts with higher-bandwidth manganite materials such as Pr$_{1-x}$Sr$_{x}$MnO$_{3}$ where the stiffness fully collapses at $T_{C}$ as expected for a second-order ferromagnetic phase transition.\cite{Baca}  Above $T$~$\approx$~200~K a spin diffusive quasielastic component develops in the low-$q$ spectral weight\cite{Lynn1996,Dai2001} arising from a short-ranged localization of electrons on the Mn$^{3+}$/Mn$^{4+}$ lattice.  This quasielastic component displays a strong field dependence\cite{Lynn1997} and dominates the spin fluctuation spectrum near $T_{C}$ with a temperature dependence that closely matches those of both lattice polarons and the bulk resistivity.\cite{Adams2000}  It is the development of this spin diffusive component, rather than the thermal population of spin waves, that truncates the ferromagnetic metallic ground state in a weakly first-order phase transition.\cite{Adams2004}

In this work we report neutron spectroscopy measurements on a single crystal sample of La$_{0.7}$Ca$_{0.3}$MnO$_{3}$.  Spin correlations at 265~K, slightly above $T_{C}$, are explored primarily through the $Q$ dependence of scattering at constant energy transfer.  The long-wavelength spin dynamics can be described by spin diffusion with a short, almost temperature-independent correlation length of $\approx$12~{\AA}.\cite{Lynn1996}  For $q$ transfers approaching the Brillouin zone edge an additional anisotropic scattering component is observed at low energies, in the form of ridges of strong quasielastic scattering intensity running along ($H$~0~0) and symmetry equivalent directions.  To the best of our knowledge, a component of this sort has not been reported in the paramagnetic phase of any other isotropic ferromagnet.  A close connection between the low-$q$ quasielastic scattering and the colossal magnetoresistance has already been established,\cite{Adams2000} supporting the possibility that the novel high-$q$ paramagnetic scattering component presented here might also play a role in the CMR physics.  In the zone boundary ($H$~$K$~0.5) plane spin correlations are observed up to energies of at least 28~meV; this scattering is qualitatively similar to the low-energy correlations in the ($H$~$K$~0) plane.

A 1.5~g single crystal sample of La$_{0.7}$Ca$_{0.3}$MnO$_{3}$, previously used for measurements of polaron correlations,\cite{Lynn2007} was grown by the floating zone technique.\cite{Shulyatev}  This is the highest composition of LCMO for which single crystal samples have been successfully grown, with $T_{C}$~=~257~K.  The crystal structure is orthorhombic perovskite, but given the presence of multiple crystallographic domains we employ the cubic notation with $a$~=~3.87~{\AA}.  The experiment was carried out at the ARCS time-of-flight chopper spectrometer at the Spallation Neutron Source, Oak Ridge National Laboratory with an incident neutron energy of 50~meV at temperatures of 100~K and 265~K. The 100~K spin wave dispersion in the ($\xi$~0~0) direction can be fit to $\hbar\omega$~=~2$S$[$|J_{1}|$(1-cos(2$\pi\xi$))+$|J_{4}|$(1-cos(4$\pi\xi$))] where the first- and fourth-nearest neighbor ferromagnetic interactions are given by $S|J_{1}|$~=~6.18~$\pm$~0.17~meV and $J_{4}/J_{1}$~=~0.19~$\pm$~0.02, roughly in agreement with the interaction values previously reported\cite{Ye} for an LCMO sample with $T_{C}$~=~238~K.
\begin{figure}
\centering
\includegraphics[width=8.6cm] {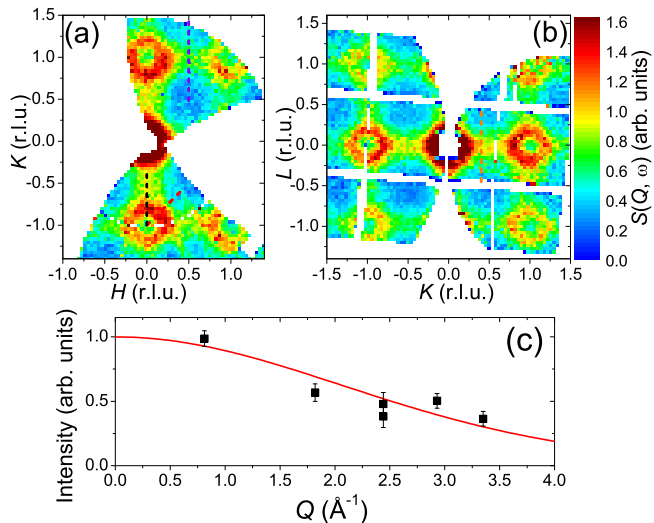} \vspace{-10mm}
\caption{(Color online) Intensity plot of $S(Q, \, \omega)$ measured at 265~K, with the energy integrated between 3~meV $\leq$~$\hbar\omega$~$\leq$~5~meV. (a) Scattering in the ($H$~$K$~0) plane.  (b) Scattering in the (0~$K$~$L$) plane.  $Q$ perpendicular to the scattering plane has been integrated over $\pm$~0.16~{\AA}$^{-1}$.  The black, red, purple, and orange dashed lines represent the scan directions displayed in Fig.~\ref{Figure2}.  (c) Integrated intensity of transverse scans through scattering ridges, with the energy integrated between 2~meV $\leq$~$\hbar\omega$~$\leq$~6~meV.  The red line is the Mn form factor squared.  Uncertainties throughout this paper are statistical and refer to one standard deviation.}\vspace{-5mm}
\label{3to5meV}
\end{figure}

Figure~\ref{3to5meV} displays an intensity plot of $S(Q, \omega)$ at $T$~=~265~K in both the ($H$~$K$~0) and (0~$K$~$L$) scattering planes with the energy transfer integrated between 3~meV $\leq$~$\hbar\omega$~$\leq$~5~meV.  This temperature is in the paramagnetic phase of LCMO, at approximately 1.03~$T_{C}$.  The most prominent features of these data are circular rings of strong scattering surrounding the Bragg positions.  These rings can be attributed to spin diffusive scattering\cite{Lynn1996} with an increase of the quasielastic linewidth as $q$ moves away from the zone center.\cite{Chatterji,DaoudAladine}  For data with energy transfers from 5~meV to at least 22~meV, the $q$ position corresponding to the maximum intensity of the ring increases with the energy value of the constant-$E$ scan consistent with the $\omega$~$\propto$~$q^{2.5}$ expectation of dynamical scaling theory;\cite{Lynn1984,Endoh,Halperin} in particular the ring radii, as measured in scans along the ($\xi$~$\xi$~0) direction, are consistent with a quasielastic half width at half maximum (HWHM) that varies with $q$ as $\Gamma(q)$~=~$\Lambda q^{2.5}$ where $\Lambda$~=~18.9~$\pm$~0.5~meV~{\AA}$^{2.5}$.  Closer to the zone boundary, the isotropy of the spin dynamics breaks down and ridges of scattering are present which connect the rings along ($H$~0~0) and equivalent directions.  These ridges of scattering are strongest in Brillouin zones at low $Q$ and the intensity falls off at higher $Q$ in a manner roughly consistent with the Mn form factor squared, as displayed in Fig.~\ref{3to5meV}(c).  These data reflect the integrated intensity of transverse scans through ridges centered at positions equivalent to (0.5~0~0), (1~0.5~0), (1.5~0~0), (1~1~0.5), (1.5~1~0), and (2~0.5~0);  all of the equivalent positions within the range of the instrument were averaged.  Given the strong coupling between the magnetic and lattice degrees of freedom in CMR manganites it is possible that these ridges of scattering also have a structural component; with the current statistics we can only note that the $Q$ dependence suggests that this scattering is primarily magnetic in origin.
\begin{figure}
\centering
\includegraphics[width=8.7cm] {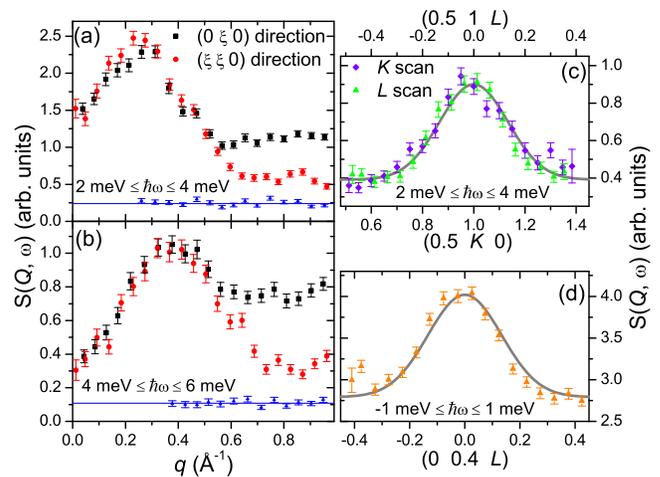} \vspace{-8mm}
\caption{(Color online) (a),(b) Differing intensities for scans in the (0~$\xi$~0) and ($\xi$~$\xi$~0) directions away from the (0~-1~0) reflection at 265~K.  The blue data points and horizontal line correspond to the background given by a scan in the ($\xi$~$\xi$~0) direction at 100~K.  (c) Scans through the (0.5~1~0) position in both directions transverse to $\vec{q}$, measured at 265~K in the $\hbar\omega$~$\approx$~3~meV data.  (d) Transverse scan through (0~0.4~0) in the elastic data, measured at 265~K.  The lines in panels (c) and (d) are Gaussian fits with a HWHM of 0.16~r.l.u. (0.26~{\AA}$^{-1}$).  For panels (a)-(c) the data have been integrated over $\pm$~0.13~{\AA}$^{-1}$ in both $\vec{q}$ directions transverse to the scan.  For panel (d) the data have been integrated over $\pm$~0.13~{\AA}$^{-1}$ along $H$ and over $\pm$~0.08~{\AA}$^{-1}$ along $K$.}\vspace{-5mm}
\label{Figure2}
\end{figure}

These ridges of additional scattering are further explored in Fig.~\ref{Figure2}, where scans in the (0~$\xi$~0) and ($\xi$~$\xi$~0) directions away from the (0~-1~0) position are shown for energies centered at 3~meV [Fig.~\ref{Figure2}(a)] and 5~meV [Fig.~\ref{Figure2}(b)].  These scans are roughly isotropic through the ring, but when $q$~$\approx$~0.55~{\AA}$^{-1}$ ($\approx$0.34~r.l.u. in the (0~$\xi$~0) direction) the data in the two directions diverge with the scattering in the ($\xi$~$\xi$~0) direction falling much faster.  The width of these ridges of scattering in directions transverse to $\vec{q}$ is about 0.26~{\AA}$^{-1}$ HWHM as shown in Fig.~\ref{Figure2}(c).  Elastic or quasielastic scattering centered at $\vec{q}$~=~(0.5~0~0) has been observed in the CMR bilayer manganite La$_{1.2}$Sr$_{1.8}$Mn$_{2}$O$_{7}$\cite{Chatterji2006} and the nearly half-doped manganite Pr$_{0.55}$(Ca$_{0.8}$Sr$_{0.2}$)$_{0.45}$MnO$_{3}$\cite{Ye2005} signifying short-range antiferromagnetic correlations.  In La$_{0.7}$Ca$_{0.3}$MnO$_{3}$, the anomalous scattering near the (0.5~0~0) position instead arises from a breakdown in dynamical scaling theory in which the energy width of the quasielastic scattering ceases to be isotropic as $q$ approaches the zone boundary, with smaller quasielastic widths for $\vec{q}$ values along the ($\xi$~0~0) direction.
\begin{figure}
\centering
\includegraphics[width=6.5cm] {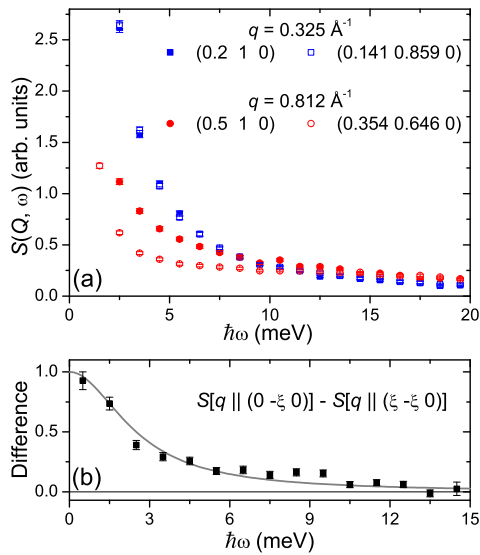} \vspace{-8mm}
\caption{(Color online) (a) Energy dependence of the 265~K paramagnetic scattering in constant-$Q$ scans.  $Q$ values have been chosen so that $\vec{q}$ away from (0~1~0) is along either the ($\xi$~0~0) or ($\xi$~-$\xi$~0) directions, and with the magnitude of $q$ either 0.325~{\AA}$^{-1}$ or 0.812~{\AA}$^{-1}$.  The data have been folded across the $K$~=~0 plane and integrated over $\pm$~0.13~{\AA}$^{-1}$ in all $\vec{q}$ directions.  (b) The energy dependence of the ridge intensity, displayed as the difference in scattering for measurements at $\vec{Q}$~=~(0~0.6~0) and (0.283~0.717~0) (both with $q$~=~0.649~{\AA}$^{-1}$).  The gray line is a guide to the eye.}\vspace{-5mm}
\label{Escan}
\end{figure}

The intensity of this scattering anisotropy is energy-dependent, as shown in the constant-$Q$ energy scans of Fig.~\ref{Escan}(a).  For smaller $q$ values where the data are well described by simple spin diffusion, such as the $q$~=~0.325~{\AA}$^{-1}$ data shown in the figure, the energy scans do not depend on the orientation of $\vec{q}$.  In the higher-$q$ data, such as the $q$~=~0.812~{\AA}$^{-1}$ data shown in the figure, the intensity of the ridge of extra scattering is demonstrated by the difference in scattering for the $\vec{q}~||~(\xi~0~0)$ and $\vec{q}$~$||$~($\xi$~-$\xi$~0) data.  Figure~\ref{Escan}(b) shows this difference in scattering intensity for $q$~=~0.649~{\AA}$^{-1}$; this $\vec{q}$ position is shifted slightly from the center of the ridge as the zone edge position will feature a nuclear superlattice reflection in the elastic data.  This anomalous scattering near the zone edge is quasielastic in nature, having a maximum at the elastic position and an energy HWHM of approximately 2.5~meV.  This anisotropy ceases to be measurable for $\hbar\omega$~$\gtrsim$~15~meV; this is roughly the energy transfer where rings of scattering surrounding adjacent Brillouin zone centers begin to overlap.  While this intensity is energy-dependent, the transverse width of the ridges is not.  A transverse scan through the ridge at (0~0.4~0) is displayed in  Fig.~\ref{Figure2}(d) for elastic scattering data (-1~meV~$\leq$~$\hbar\omega$~$\leq$~1~meV), showing a width (HWHM of 0.26~{\AA}$^{-1}$) equal to that observed in the inelastic data.
\begin{figure}
\centering
\includegraphics[width=8.6cm] {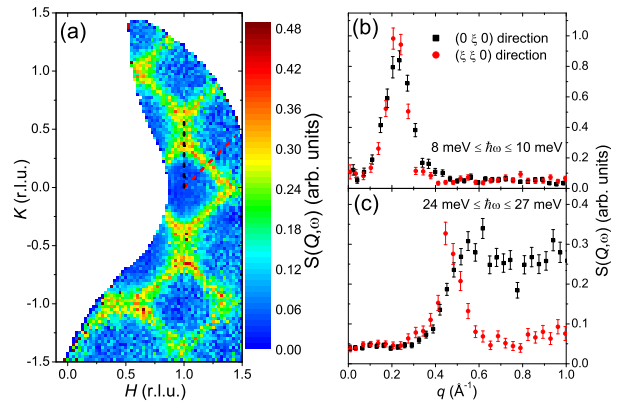} \vspace{-6mm}
\caption{(Color online) (a) Intensity plot of $S(Q, \, \omega)$ measured at 100~K, with the energy integrated between 24~meV $\leq$~$\hbar\omega$~$\leq$~27~meV.  $Q$ perpendicular to the scattering plane has been integrated over $\pm$~0.16~{\AA}$^{-1}$.  The black and red dashed lines represent the scan directions in the other panels.  (b)-(c)  Differing intensities for scans in the (0~$\xi$~0) and ($\xi$~$\xi$~0) directions away from the (1~0~0) reflection at 100~K.  The data have been integrated over $\pm$~0.13~{\AA}$^{-1}$ in both $\vec{q}$ directions transverse to the scan.}\vspace{-5mm}
\label{100K}
\end{figure}

These spin correlations above $T_{C}$ can be compared to the propagating spin waves observed below $T_{C}$.  Figure~\ref{100K} displays data measured at 100~K.  For relatively low energies, such as $\hbar\omega$~$\approx$~9~meV as shown in Fig.~\ref{100K}(b), the spin waves are isotropic.  At 100~K we do not observe any low-energy anisotropy in $S(\vec{Q}, \, \omega)$, in contrast to the anomalous quasielastic ridges of scattering along the ($H$~0~0) direction observed in the paramagnetic phase.  This is consistent with previous reports\cite{Lynn1996} of the low-$q$ quasielastic scattering developing only at temperatures approaching $T_{C}$.  As expected, the widths of these spin waves in a constant-$E$ scan (a HWHM of about 0.06~{\AA}$^{-1}$) are determined by the instrumental resolution and are far narrower than the $Q$-space peaks arising in the data above $T_{C}$.  At higher energies, such as  $\hbar\omega$~$\approx$~25.5~meV shown in Figs.~\ref{100K}(a) and \ref{100K}(c), the peak in $Q$ space approaches the zone boundary in the ($H$~0~0) direction and thus the isotropy breaks down.  The constant-$E$ correlations in the ($H$~$K$~0) plane develop into square shapes which merge near the zone boundaries, consistent with the calculated spin wave scattering.
\begin{figure}
\centering
\includegraphics[width=8.5cm] {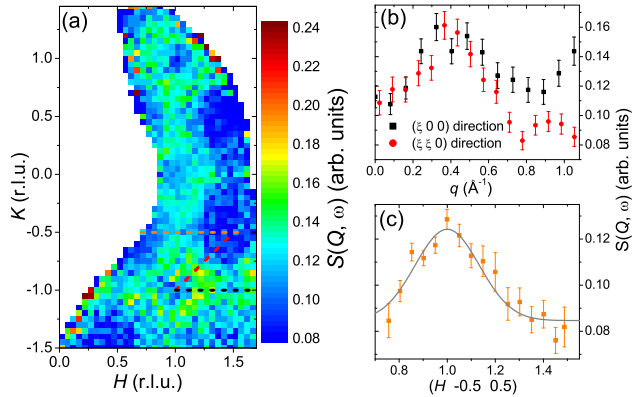} \vspace{-6mm}
\caption{(Color online) Scattering intensities with the energy integrated between 24~meV $\leq$~$\hbar\omega$~$\leq$~28~meV and $|L|$~=~0.5, measured at 265~K.  (a) Intensity plot of the ($H$~$K$~0.5) scattering plane.  $Q$ perpendicular to the scattering plane has been integrated over $\pm$~0.16~{\AA}$^{-1}$.  The black, red, and orange dashed lines represent the scan directions in the other panels.  (b) Differing intensities for scans in the ($\xi$~0~0) and ($\xi$~$\xi$~0) directions away from the (1~-1~0.5) reflection.  (c) Scan in the ($\xi$~0~0) direction through the (1~-0.5~0.5) position.  The Gaussian fit has a HWHM of 0.16~r.l.u. (0.26~{\AA}$^{-1}$).  The data in panels b and c have been integrated over $\pm$~0.13~{\AA}$^{-1}$ in both $\vec{q}$ directions transverse to the scan.}\vspace{-5mm}
\label{HighE}
\end{figure}

The presence of persistent spin correlations at 265~K can be further explored through the spin dynamics in the ($H$~$K$~0.5) scattering plane.  For low energy transfers, the spin correlations in this plane will present as scattering centered at integer positions, arising from the ridges of scattering displayed in Fig.~\ref{3to5meV}.  For higher energy transfers, such as 24~meV~$\leq$~$\hbar\omega$~$\leq$~28~meV for the data shown in Fig.~\ref{HighE}, the magnetic scattering displays a more complex structure.  To increase statistics the measured data have been folded across the $L$~=~0 plane, so that data with $L$~=~-0.5 and $L$~=~0.5 have been averaged together.  The intensity plot of correlations in the ($H$~$K$~0.5) plane, shown as Fig.~\ref{HighE}(a), is qualitatively quite similar to the correlations in the ($H$~$K$~0) plane at lower energies.  A similar anisotropy between scans along the ($\xi$~0~0) and ($\xi$~$\xi$~0) directions is observed, as shown in Fig.~\ref{HighE}(b); these ridges have a transverse width, shown in Fig.~\ref{HighE}(c), consistent with the low-energy data.  It should be noted that the ($H$~$K$~0.5) scattering plane lies along a Brillouin zone edge, so that all correlations in this plane are in the high-$q$ regime where simple spin diffusion theory should not be applicable.  In particular, the correlations yield peaks in $q$ that are far narrower than would be expected from a purely diffusive model.  A breakdown in scaling theory in which the width of peaks in constant-$E$ scans falls far below the theoretical value at high $q$ was also observed in paramagnetic iron and nickel.\cite{Lynn1984}  These correlations are also qualitatively similar to the ferromagnetic phase $Q$-space spin wave correlations displayed in LCMO.  The bilayer manganite La$_{1.2}$Sr$_{1.8}$MnO$_{7}$ was likewise reported\cite{Chatterji} to display $Q$-space spin correlations in the paramagnetic phase that qualitatively resembled those of ferromagnetic spin waves.  Given the dispersion relation displayed by propagating spin waves in LCMO below $T_{C}$, the $Q$-space correlations in the ($H$~$K$~0) plane at an energy of $\omega_{0}$ will have the same structure as correlations in the ($H$~$K$~0.5) plane at an energy of $\omega_{0}$+$\Delta\omega$ where $\Delta\omega$~=~$4S|J_{1}|$ (such that $\Delta\omega$~$\approx$~25~meV at 100~K).  Finding the correlations of Fig.~\ref{HighE} in data where $\hbar\omega$~$\approx$~26~meV suggests a significant renormalization of $\Delta\omega$ from the 100~K data; this is reminiscent of the previously reported renormalization of the spin wave stiffness,\cite{Lynn1996,Lynn1997} where $D(T_{C})$~$\approx$~$D(T=0)$/2.

It is clear that the spin correlations in LCMO above $T_{C}$ result in well-defined peaks in constant-$E$ scans that qualitatively resemble the correlations from the spin wave excitations below $T_{C}$; similar results have been previously reported in a bilayer manganite.\cite{Chatterji}  Despite these well-defined peaks in constant-$E$ scans, no clear peaks at finite energy are observed in constant-$Q$ scans.  For $\vec{q}$ positions away from the ($H$~0~0) direction, the positions of these peaks are well described by spin diffusive dynamical scaling theory.  The data at higher-$q$ values deviate from the expectations of dynamical scaling theory primarily through peaks in the $Q$-space correlations that are far narrower than the simple spin diffusive model would predict.  An additional component of the paramagnetic scattering is also observed as ridges of unexpectedly strong quasielastic scattering at low energy transfers and $\vec{q}$ positions parallel to $\vec{a}^{\ast}$ or a symmetry equivalent direction.  Intrinsic inhomogeneity with small scale regions of competing phases is a common signature in strongly correlated electron materials, including stripe order in high-$T_{C}$ superconducting cuprates\cite{Tranquada} and polar nanoregions in relaxor ferroelectrics.\cite{Blinc}  Phase separation of this sort is well known in CMR manganites, with considerable evidence for lattice polarons\cite{Lynn2007} and spin polarons;\cite{deTeresa}  LCMO samples with smaller doping levels also display evidence of ferromagnetic droplets.\cite{Hennion1998}  The physics of colossal magnetoresistance in LCMO has been modeled as a percolation\cite{Mayr} or Griffiths phase\cite{Salamon} effect arising from the separation of various competing phases.

The effects of hole doping and applied field on the spin fluctuation spectrum of LCMO near $T_{C}$ suggest that the small-$q$ spin diffusive portion of the scattering arises from the short length-scale hopping of electrons on the Mn$^{3+}$/Mn$^{4+}$ lattice;\cite{Lynn1996} this diffusive scattering coexists with spin waves near $T_{C}$ and drives the ferromagnetic phase transition.  It is also known that the temperature dependence of the low-$q$ quasielastic scattering is quite similar to the temperature dependences of the bulk resistivity and the polaron correlations, suggesting a close connection between paramagnetic scattering and colossal magnetoresistance.  The new high-$q$ quasielastic scattering in the paramagnetic phase described in this work represents spin correlations with wavelengths approaching atomic length scales, comparable in size to the small polarons generated by localized electrons.  We hope that further measurements on the short-range spin correlations near $T_{C}$ will shed new light on the physics of colossal magnetoresistance.

J.S.H. acknowledges support from the NRC/NIST Postdoctoral Associateship Program.  This research at Oak Ridge National Laboratory's Spallation Neutron Source was sponsored by the Scientific User Facilities Division, Office of Basic Energy Sciences, U. S. Department of Energy.\\
\\
$\ast$ email: joel.helton@nist.gov \\
$\dag$ email: jeffrey.lynn@nist.gov
\bibliography{LCMO}
\end{document}